\documentclass[10pt,aps,prl,twocolumn,showpacs,superscriptaddress]{revtex4-1} 

\usepackage[utf8]{inputenc}
\usepackage{graphicx}

\usepackage{amsmath}
\usepackage{amssymb}
\usepackage{bm}
\usepackage{txfonts}

\usepackage{multirow}
\usepackage{mathptmx} 
\usepackage{siunitx}

\usepackage{xargs}                    
\usepackage[pdftex,dvipsnames]{xcolor}

\usepackage[%
  colorlinks=true,
  urlcolor=blue,
  linkcolor=blue,
  citecolor=blue
]{hyperref}

\begin{document}

\title{Measuring ion oscillations at the quantum level with fluorescence light}

\author{G.~Cerchiari}
\email{giovanni.cerchiari@uibk.ac.at}
\affiliation{Institut f\"ur Experimentalphysik, Universit\"at Innsbruck, Technikerstrasse~25, 6020~Innsbruck, Austria}
\altaffiliation{Corresponding author}
\author{G.~Araneda}
\affiliation{Institut f\"ur Experimentalphysik, Universit\"at Innsbruck, Technikerstrasse~25, 6020~Innsbruck, Austria}
\affiliation{Department of Physics, University of Oxford, Clarendon Laboratory, Parks Road, Oxford OX1 3PU, U.K}
\author{L.~Podhora}
\affiliation{Department of Optics, Palack\'y University, 17. Listopadu 12, 77146 Olomouc, Czech Republic}
\author{L.~Slodi\v{c}ka}
\affiliation{Department of Optics, Palack\'y University, 17. Listopadu 12, 77146 Olomouc, Czech Republic}
\author{Y.~Colombe}
\affiliation{Institut f\"ur Experimentalphysik, Universit\"at Innsbruck, Technikerstrasse~25, 6020~Innsbruck, Austria}
\author{R.~Blatt}
\affiliation{Institut f\"ur Experimentalphysik, Universit\"at Innsbruck, Technikerstrasse~25, 6020~Innsbruck, Austria}
\affiliation{Institut f\"ur Quantenoptik und Quanteninformation, \"Osterreichische Akademie der Wissenschaften, Technikerstrasse 21a, 6020 Innsbruck, Austria}

\date{\today}

\begin{abstract}

We demonstrate an optical method for detecting the mechanical oscillations of an atom with single-phonon sensitivity. The measurement signal results from the interference between the light scattered by a single trapped atomic ion and that of its mirror image. The motion of the atom modulates the interference path length and hence the photon detection rate. We detect the oscillations of the atom in the Doppler cooling limit and reconstruct average trajectories in phase space. We demonstrate single-phonon sensitivity near the ground state of motion after EIT cooling. These results could be applied for motion detection of other light scatterers of fundamental interest, such as trapped nanoparticles.
\end{abstract}
\maketitle

Observing trapped ion oscillations is a well established research technique for precision measurements of fundamental constants and studying fundamental quantum physics~\cite{Eliseev2015direct, FundamentalPhysicsinParticleTraps, Leibfried1996}. In the trap, ions are confined in free space and are well isolated from the surrounding environment. Observation of their oscillations around the equilibrium position allows, for example, the precise determination of the masses of atomic elements and of the electron~\cite{Block2016, Sturm2011g}, as well as testing the symmetries between matter and antimatter~\cite{Smorra2017partsperbillion, Ulmer2015}. Motional quantum effects can also be explored and manifest themselves at the lowest temperatures of the oscillator~\cite{Niemann2019, biercuk2010ultrasensitive, jost2009entangled}. In trapped ions, cooling and detection of non-classical states of motion are achieved via laser interaction~\cite{Monroe1996, Leibfried1997}. These states are observed by reconstructing the Wigner function by quantum state tomography~\cite{Poyatos1996} or by measuring phonon state populations~\cite{Meekhof1996}. In such experiments, sensitivity at the quantum level relies on the electronic structure of the ion, which allows resolving and addressing the motional sidebands in one of its transitions~\cite{leibfried2003quantum}.

In larger systems than ions, such as mesoscopic micromembranes, microcantilevers and levitated nanoparticles, the measurement of oscillations offers interesting prospects for force sensing and fundamental investigations in the quantum regime~\cite{aspelmeyer2012quantum, ranjit2016zeptonewton}. Unlike atoms, these systems do not possess narrow internal transitions, but they can be cooled to the motional ground state~\cite{o2010quantum, teufel2011sideband, chan2011laser}. As for atomic ions, levitated nanoparticles are trapped in free space. For nanoparticles, the lowest temperatures are achieved via coherent light scattering in a Fabry--Perot cavity~\cite{delic2019cavity,windey2019cavity,deli2019motional}. Another approach is the direct observation of the particle position via light scattering for active feedback cooling~\cite{Tebbenjohanns2019experiment}. In this regime, resolving single quanta of motion should allow feedback cooling to the ground state and the observation of quantum states of motion.

Experiments with atomic ions have used the light scattered on a dipole transition for motion analysis and feedback cooling~\cite{Bushev2006, Bushev2013, slodivcka2012interferometric, rotter2008monitoring}, however sensitivity at the single quantum of motion has remained elusive. Recently, it has been predicted that the light scattered by a dipole emitter could grant such sensitivity~\cite{Tebbenjohanns2019}. Thus, feedback cooling of a nanoparticle to its ground state of motion should be possible, without the need for an optical cavity. Similarly, measurement of atomic motion at the quantum level may be accessible without the need for sideband spectroscopy on a narrow transition. In this Letter, we demonstrate that self-interference of light scattered by a single trapped ion can provide sensitivity at the single-phonon level. First, we show simultaneous detection of all motional modes of the ion. Then, we describe the reconstruction of phase space trajectories. Finally, we investigate the limit of this technique for the detection of motion near the ground state of the mechanical oscillator.

\begin{figure}
\centerline{\includegraphics[width=1.0\columnwidth]{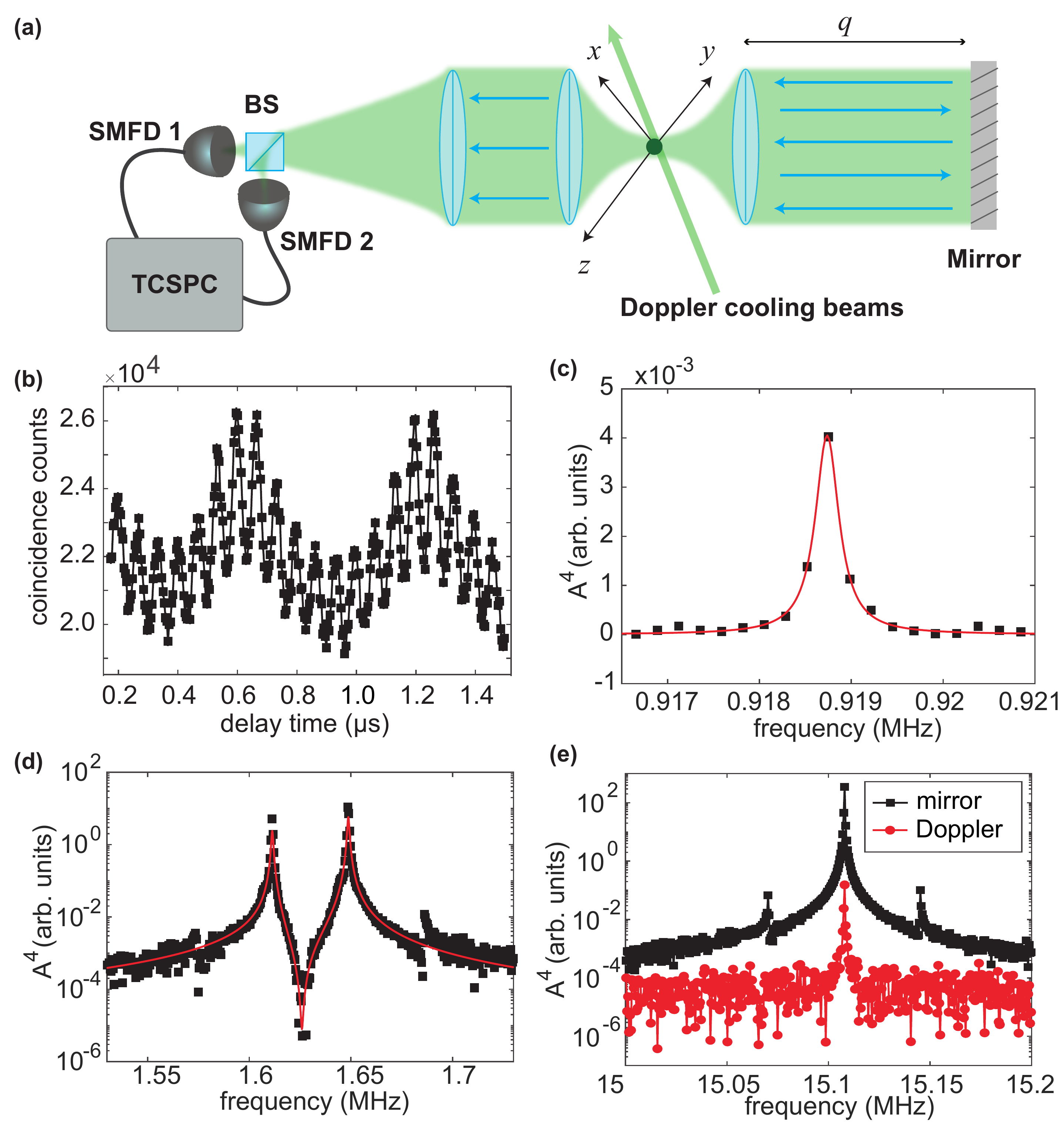}}
\caption{Detection of the ion motion by self-interference of the light emitted during continuous Doppler cooling. (a) Optical setup. (b) Autocorrelation histogram of the photon detection events. Micromotion is visible, superimposed to a slower secular oscillation in the radial modes ($x$ and $y$ axes). The line is meant to guide the eye. (c-e) Power spectra of the autocorrelation. (c) Oscillation on the axial mode ($z$ axis), orthogonal to the optical axis. (d) Radial modes. The red curves in (c, d) are fit functions to the data. (e) Micromotion. The black data corresponds to the setup (a), while the red data are taken while blocking the reflection of the mirror. The lines are meant to guide the eye.}
\label{fig:setup}
\end{figure}

We confine a single $^{138}$Ba$^+$ ion in a linear Paul trap. The motion of the ion has three orthogonal modes of oscillation with frequencies $\omega_x/ 2\pi \sim 1.61 $~MHz, $\omega_y/ 2\pi \sim 1.65 $~MHz and $\omega_z/ 2\pi \sim 0.92 $~MHz. The radiofrequency field used for radial confinement is driven at $\Omega_\text{rf}/ 2\pi \sim$ $15.1$~MHz. The ion is Doppler cooled on the $6\textrm{S}_{1/2} \leftrightarrow 6\textrm{P}_{3/2}$ transition at 493\,nm with a laser beam propagating in a direction overlapping with all three normal modes~\cite{araneda2019thesis}. The motion of the ion is studied by detecting the fluorescence light at 493\,nm emitted by the ion in the interferometric configuration depicted in Fig. \ref{fig:setup}(a). The fluorescence light is imaged by two in-vacuum confocal objectives with $\text{NA} = 0.40$~\cite{gerber2009quantum}. The light collected by one of the objectives is reflected by a distant mirror ($\sim 30\,$cm away from the ion) back to the ion. The reflection is superimposed to the direct fluorescence through the other objective. The photons collimated by the second objective are directed to two single-mode fiber avalanche photodiode (APD) detectors using a beam splitter. The arrival of each photon is time-tagged with a two-channel Time-Correlated Single Photon Counting system (TCSPC, PicoHarp 300). The TCSPC has 4~ps resolution. The experiments are limited both by the dead-time ($100$~ns) and by the jitter ($0.8$~ns) of each APD.
The photon count rate $R$ results from the interference of the primary and reflected light beams. The measured rate depends on the mutual distance $q$ between the ion and the mirror through
\begin{equation} \label{eq_fringe}
    R = R_0 \Bigg(1 + \mathcal{V} \sin\left(\frac{4\pi}{\lambda}q\right)\Bigg) \,,
\end{equation}
where $\mathcal{V}$ is the contrast of the interference and $R_0$ is the average photon rate. If the position of the mirror is kept fixed, $q$ depends only on the ion oscillations resulting in a modulation of the photon rate $R$. Real-time measurement of $q$ was demonstrated in Ref.~\cite{Bushev2006}. The scheme required narrow-bandwidth amplification by mixing the photon stream with a local oscillator signal. The interference of scattered light is already used for detection of nanoparticle motion~\cite{Tebbenjohanns2019experiment}. However, the interference is typically achieved homodyning the focused Gaussian beam illuminating the nanoparticle with the scattered dipole field. Our scheme based on self-interference is more flexible by allowing orientation of the detection at an arbitrary angle. Moreover, we collect the position information contained in both the primary and reflected fluorescence light that interfere on the detector. This two features can be used to realize an ideal theoretical arrangement~\cite{Tebbenjohanns2019} that should provide detection at the Heisenberg limit (see Suppl. Material D).

In this work, photon arrival times are recorded and post-processed. Time-tagging allows detecting all the modes simultaneously without band filtering. Motion detection using photon time-tagging and post-processing was demonstrated in Ref.~\cite{dholakia1993photon}, where, however, the photon rate was solely modulated by the Doppler effect without the use of optical interference.

Continuous Doppler cooling is applied by the 493~nm beam detuned by $\Delta/ 2\pi = -35$~MHz and with a Rabi frequency $\Omega/ 2\pi \sim 12 $~MHz. The repumper at 650~nm is detuned of $\Delta_r/ 2\pi = -10$~MHz and induces a Rabi frequency of $\Omega_r/ 2\pi \sim 14.2 $~MHz. These values are estimated by fitting Bloch equation steady-state solutions to scans of the fluorescence intensity~\cite{schubert1995transient}. In this condition, a visibility $\mathcal{V} \sim 33-37\%$ is typically measured by scanning the mirror position over an interference fringe with a global detected photon rate of $\left(5-6\right)\times10^4$~s$^{-1}$. To detect the motion, photon events are registered for $\sim 900$~s under equilibrium condition, i.e. under steady state Doppler cooling of the trapped ion. Figure~\ref{fig:setup} presents the analysis of data acquired in a single experimental run. The ion motion is studied by computing the autocorrelation of the photon events. A section of the autocorrelation is shown in Fig.~\ref{fig:setup}(b) where both secular and micromotion oscillations are visible. Fig.~\ref{fig:setup}(c)-(e) show the frequency peaks present in the power spectrum of the autocorrelation. Since the autocorrelation function is the Fourier transform of the power spectrum, further calculating the power spectrum of the autocorrelation leads to a spectral signal proportional to the fourth power of the excursion amplitude $A$ (see scales in the Fig.~\ref{fig:setup}). The peak in Fig.~\ref{fig:setup}(c) corresponds to the axial (z axis) mode, which has the smallest overlap with the optical axis of the confocal objectives. Fig.~\ref{fig:setup}(d) shows the x and y secular modes. The system can be modelled assuming that the ion evolves following free trajectories of oscillation on the three normal modes randomly perturbed at discrete times. Such events are generated by photon absorption and emission, collisions with the buffer gas or perturbation in the trapping electromagnetic field. All these events modify the oscillator phase with a random re-initialization, but affect all the modes simultaneously. We used these assumptions to calculate suitable fit functions (red curves in Fig.~\ref{fig:setup}(c,d)) for our data in linear expansion of Eq.~\ref{eq_fringe} in $q/\lambda$. Our model describes the data for 5 orders of magnitude near the oscillation peaks of the radial modes (x and y), also identifing the interference minimum that appears at $\sim1.625$~MHz between the modes (see Fig.~\ref{fig:setup}d). The minimum arises from the analysis of the motion via the two-photon correlation function, which fixes the phases of oscillations of the different modes to zero at correlation time $t=0$. Further non-linear terms in the expansion series of Eq.~\ref{eq_fringe} contribute with additional peaks in the spectrum. Thanks to the non-linear peaks at $2\omega_x-\omega_y\sim1.57$~MHz and $2\omega_y-\omega_x\sim1.69$~MHz we estimate that the ion oscillates with 115--125~nm excursion in the radial modes (peak-to-peak). More details about the model function and the amplitude estimation can be found in Suppl. Material section B. Fig.~\ref{fig:setup}(e) shows the micromotion peak. Micromotion can also be detected without the reflection from the mirror (red data points) thanks to the Doppler effect~\cite{bluemel1989chaos}. However, our measurement shows that the self-interference technique (black data points) delivers a 5-7 fold improvement in amplitude sensitivity \cite{explain1}. 

Ion motion analysis is not restricted to ergodic conditions, but the motion can also be detected in reference to a state preparation. In this way, time-tagged photon detection events can also be used to reconstruct the trajectory of the oscillator in phase space. Similarly to other schemes~\cite{Sturm2011phase, Eliseev2013phase}, we impress a phase to the ion oscillation by and external drive. The position and phase of the harmonic motion is obtained by synchronizing time tagging with the drive phase and by averaging over several realizations.
\begin{figure}
\centering
\includegraphics[width=1\columnwidth]{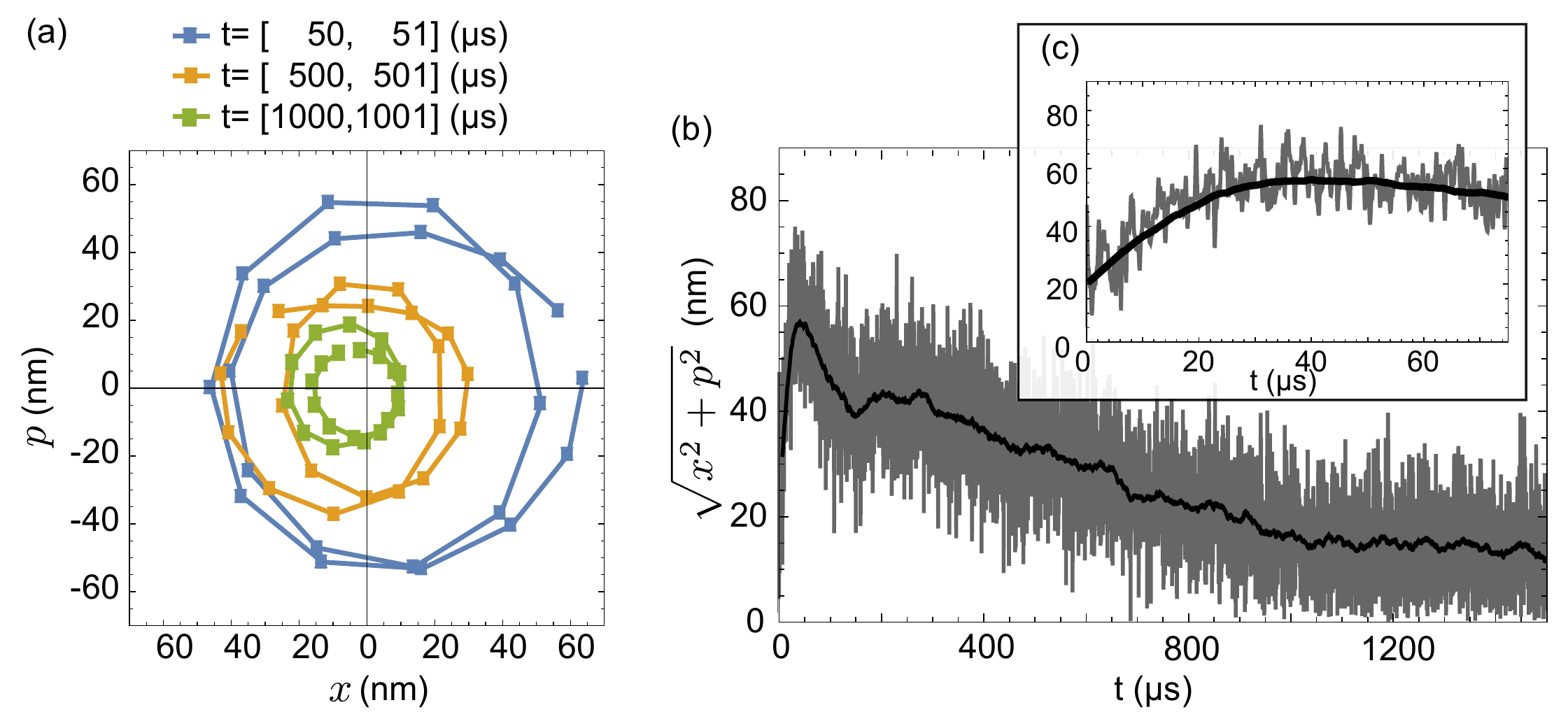}
\caption{Ion motion after an initial coherent excitation. (a) Average phase space trajectory in three $1\,$\si{\micro}s-long time windows at $t = \{50,\,500,\,1000\}\,$\si{\micro}s. The slight asymmetry of the trajectories is caused by shot noise in the measured oscillation amplitude, see Suppl. Material section C. Error bars ($\sim$15~nm in both quadratures) are omitted for readability. (b) Phase space amplitude of the coherent oscillation. The gray line shows the original data; the black line is a 8~\si{\micro}s moving average. The oscillation profile is displayed up to 1.4~ms, including the rf-driven part and ringdown. (c) shows a zoom on the first 80~\si{\micro}s.}
\label{fig:phase}
\end{figure}
In the experiment, the trapped ion is driven on its y-mode with an external rf electric field while being Doppler cooled with the same parameters as described above. The drive field is applied for 50~\si{\micro}s, which results in 100~nm excursion oscillation with a well-defined phase. The experiment is repeated $\sim 5\cdot10^5$ times. The starting point of the rf drive is used for correlation analysis to calculate the histogram of photon events arriving at specific time delays from the drive start. The histogram is filtered and analyzed in Fourier space to reconstruct the experimental trajectories of the position $q$ and momentum $p$ quadratures (see Suppl. Material C1). The $p$~quadrature is normalized by a factor $1/m\omega_y$, where $m$ is the mass of the ion. The combined position and momentum data are plotted in phase space in Fig.~\ref{fig:phase}(a) for different time intervals, showing features of a coherent state with decaying amplitude. Fig.~\ref{fig:phase}(b) shows the amplitude of the oscillation calculated as the distance from the center of the phase space. An exponential decay fit to the $q$-quadrature gives a $\tau = 847(12)$~\si{\micro}s time constant. The coherence time is unchanged when using a pulsed, instead of continuous, illumination, ruling out an effect of the cooling laser. Further measurements by sideband spectroscopy indicate a trap heating rate $<10$~phonon/s (see below). Thus, we concluded that the observed decay is due to progressive dephasing of the ion oscillations caused by instability in the rf trap drive (see Suppl. Material A2). 

\begin{figure}
\centerline{\includegraphics[width=1.0\columnwidth]{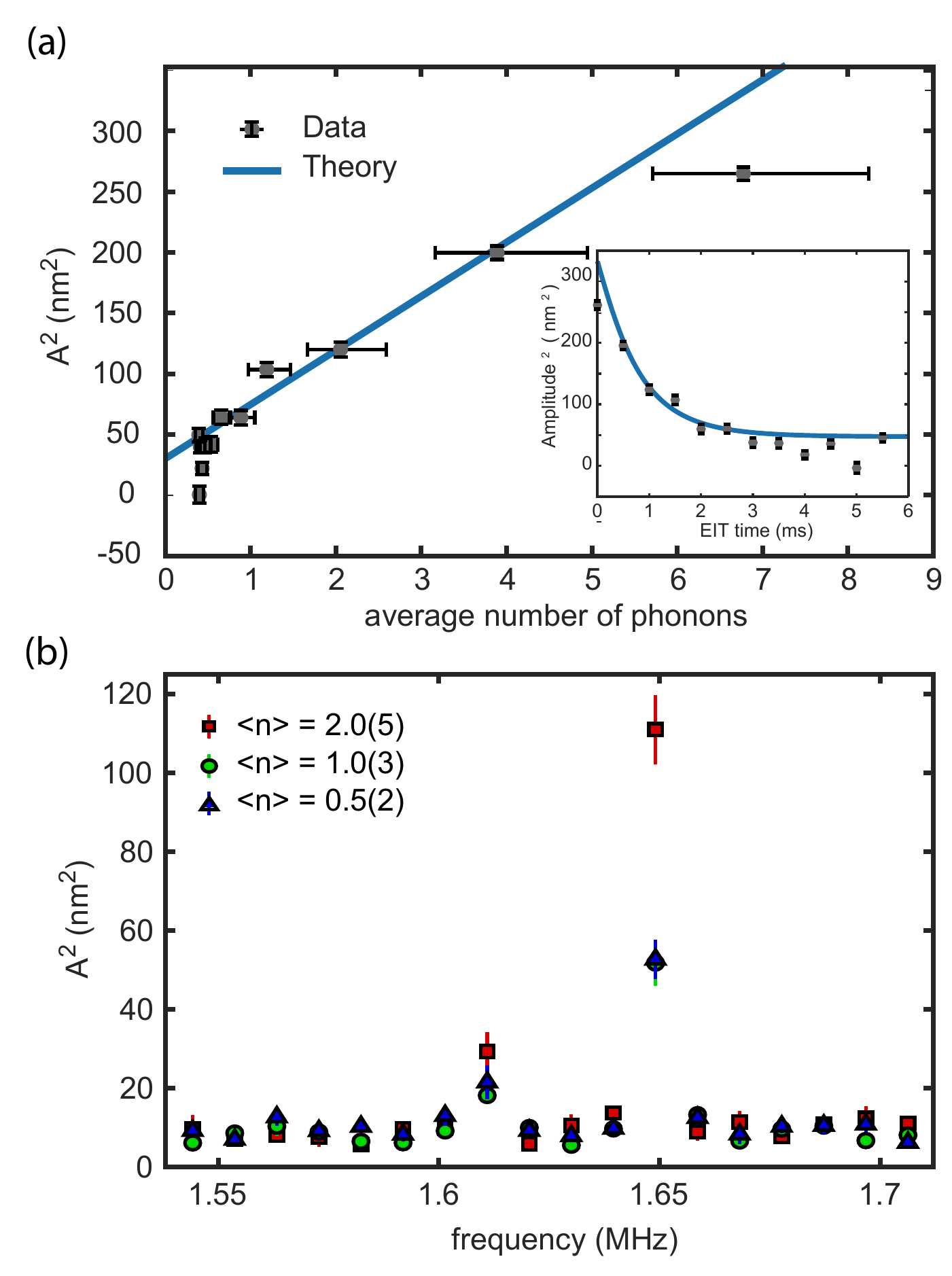}}
\caption{(a) Power spectral density of the $\omega_y/2\pi=1.65$~MHz mode as function of the number of phonons. (Inset) Same data shown as a function of the EIT cooling time. The blue lines correspond to the theoretical value of the motional wavepacket size calculated for $^{138}$Ba$^+$. (b) Power spectrum of the radial ion motion for initial phonon numbers 2.0(5), 1.0(3), and 0.5(2). The spectra and the uncertainty of the data points are calculated averaging five measurements of 900~s acquired as described in the text.
}
\label{fig:EIT_cooling_amplitude}
\end{figure}

To study the sensitivity of this method near the quantum ground state of motion, the radial modes of a single ion are cooled using electromagnetically induced transparency (EIT; see Suppl. Material section A1)~\cite{schmidt2001laser}. The average phonon number $\langle n \rangle$ is measured by sideband spectroscopy on the $6\textrm{S}_{1/2} \rightarrow 5\textrm{D}_{5/2}$ transition at 1.7~\si{\micro}m by comparing the strengths of the red and blue sidebands~\cite{leibfried2003quantum}. We vary the length of the EIT cooling pulse between 0.1 to 5 ms to obtain $\langle n \rangle$ between 7 and 0.5. After EIT cooling, the motion of the ion is probed using the Doppler beam for 120~\si{\micro}s, which causes heating of the ion motion at a linear rate $\sim$ 0.12 phonons / (100~\si{\micro}s), measured by sideband spectroscopy. The sequence consisting of Doppler cooling, EIT cooling, and pulsed Doppler illumination is repeated $\sim 6\cdot10^5$ times over 900~s, during which the setup is sufficiently stable (Suppl. Material section A3). Figure~\ref{fig:EIT_cooling_amplitude} shows detection of the ion motion at low phonon number using the autocorrelation signal of the photons scattered from the Doppler beam. The spectral power at $\omega_y$ as a function of the EIT cooling time is given in Fig.~\ref{fig:EIT_cooling_amplitude}(a). Here, the statistical uncertainties of the spectral powers are estimated from the background in the spectrum. The blue curve in the main figure is the theoretical value of the squared wave packet size, given by $(\hbar/m\omega_y)(\langle n\rangle+1/2) $. The measured points are offset and scaled in the vertical axis to calibrate the amplitude such that the point at $\langle n \rangle = 2$ corresponds to the theoretical value. This data allows preparing the initial phonon numbers $\langle n \rangle = 2.0(5)$, 1.0(3), and 0.5(2) with suitable EIT pulse lengths. For these particular values, the measurement is repeated five times. These acquisitions are used to compute the power spectra of Fig.~\ref{fig:EIT_cooling_amplitude}(b) evaluated from the photon autocorrelation. Considering the $\omega_y$ mode selected for calibration, the two-phonon and one-phonon spectral powers are distinguished by a 5.6~$\sigma$ deviation. At phonon number 1.0(3) and 0.5(2), our analysis indicates that the two measurements are compatible ($<1\sigma$). We attribute this to instabilities in the interference fringe offset over the total measuring time of several hours (see Suppl. Material section A3). The measurements presented in Fig.~\ref{fig:EIT_cooling_amplitude} show how our method can be applied to study cooling and heating dynamics of diverse mechanical oscillators for multiple motional modes simultaneously.

In this Letter, we have demonstrated that the single photons emitted by a trapped ion can be analyzed to detect its motion at the single quantum level. We have presented direct optical detection of the ion's mechanical oscillations and measured their amplitude, frequency and phase. The method could be used to detect non-classical states of motion, such as superposition of Fock or Coherent states, in a full-optical manner. Fock states, for example, could be detected by comparing the spectral information of the signal acquired on the slope of the interference fringe to access $\langle q \rangle$, and near the dark fringe to measure $\langle q^2 \rangle$. 
The presented methods do not rely on the internal electronic structure of the ion and, thus, may find application for other oscillators such as levitated nanoparticles. In typical setups, the light field scattered by the nanoparticle is interfered with the illumination beam in the forward direction~\cite{Tebbenjohanns2019}. For oscillations along the direction orthogonal to the beam and its polarization axis, our configuration can improve the sensitivity to motion by more than one order of magnitude compared to the forward detection scheme (Suppl. Material section D).

With the presented techniques, all oscillation modes may be analyzed simultaneously. In a trapped ion chain, a chosen ion could be monitored to detect all the frequency modes of the ion string. Thus, our technique may find application for sympathetic cooling and study of oscillations of ion species that cannot be directly laser cooled~\cite{sheridan2011weighing}. Furthermore, the scheme can be used for detecting the motion of ions in a Penning trap, by adapting the trap for optical access such as realized in Ref.~\cite{Kellerbauer2014}. These extensions can find applications in fundamental physics beyond the field of quantum optics.

\hfill \break
\textit{Acknowledgements.} We thank P. Bushev, L. Dania and A. Rischka for insightful discussions and J. Braun for proofreading the manuscript. L. P. is grateful for Grant No. CZ.02.2.69/0.0/0.0/16\_027/0008482 of MEYS CR and grant No. GA19-14988S of the Czech Science Foundation. This work was supported by the European Commission through project PIEDMONS 801285 and the Marie Skłodowska-Curie Action, Grant Number: No 801110 (Erwin Schrödinger Quantum Fellowship Programme). This work was also supported by the Institut für Quanteninformation GmbH.
\hfill \break

\bibliography{bibliography_deltan1}

\clearpage

\thispagestyle{empty}
\onecolumngrid
\begin{center}
\vspace{5 mm}
\textbf{\large Supplementary Information for:\\Measuring ion oscillations at the quantum level with fluorescence light}\\
\vspace{2 mm}
G. Cerchiari,$^1$, G. Araneda$^{1,2}$, L. Podhora$^{3}$, L. Slodi\v{c}ka$^{3}$, Y. Colombe,$^1$ and R. Blatt,$^{1,4}$ \\
\vspace{2 mm}

\textit{\small
$^1$Institut f\"{u}r Experimentalphysik, Universit\"{a}t Innsbruck, Technikerstra\ss e 25, 6020 Innsbruck, Austria\\
$^2$Department of Physics, University of Oxford, Clarendon Laboratory, Parks Road, Oxford OX1 3PU, U.K\\
$^3$Department of Optics, Palack\'y University, 17. Listopadu 12, 77146 Olomouc, Czech Republic\\
$^4$Institut f\"{u}r Quantenoptik und Quanteninformation, \"{O}sterreichische Akademie der Wissenschaften, Technikerstra\ss e 21a, 6020 Innsbruck, Austria\\
}

\vspace{10 mm}
\end{center}
\twocolumngrid

\section{A: Setup details}

\subsection{A1: EIT cooling setup}\label{sec:EITcooling}
\begin{figure}[h]
\centering
\includegraphics[width=1\columnwidth]{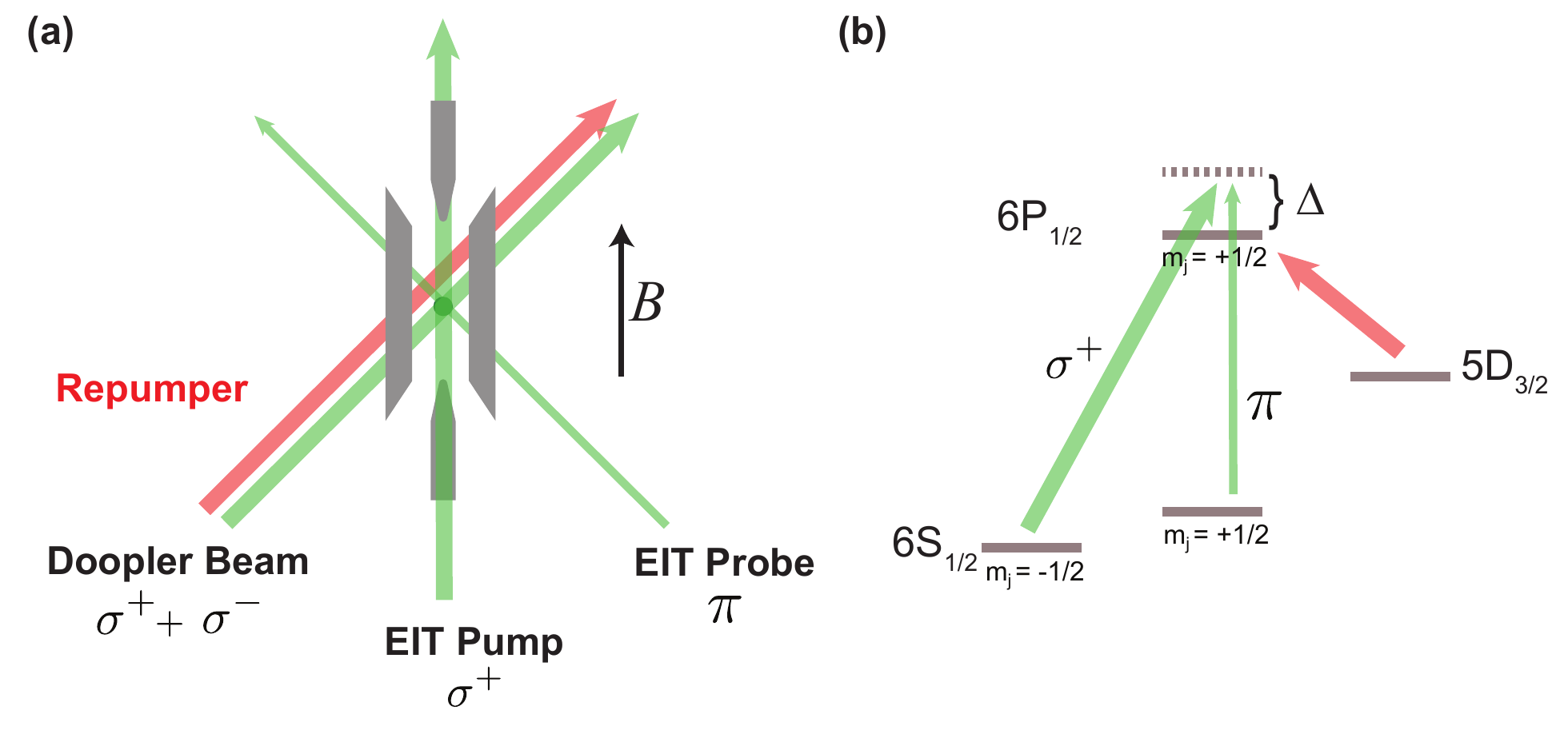}
\caption{Laser configuration used for EIT cooling. (a) Schematic configuration of the light beams in the setup, (b) Simplified level structure of $^{138}$Ba$^+$ with laser detunings. The thickness of the arrows (not to scale) suggests the intensity relation of the different beams.}
\label{fig:EIT_setup}
\end{figure}

We implemented Electromagnetic Induced Transparency (EIT) cooling by addressing the $6\textrm{S}_{1/2} \leftrightarrow 6\textrm{P}_{3/2}$ transition at 493~nm. The geometry of the setup and a simplified level scheme for $^{138}$Ba$^+$ are schematically presented in Fig.~\ref{fig:EIT_setup}. Our realization takes advantage of an already predisposed magnetic field $\vec{B}$ which is aimed to define a quantization axis inside the apparatus. For EIT cooling, two laser beams are necessary: a ``pump" beam and a ``probe" beam. The pump beam propagates parallel to the magnetic field with circular-right polarization, driving the $\sigma^+$ transition. The probe beam propagates with an angle of 45$^\circ$ with respect to the magnetic field, with horizontal polarization to drive $\pi$-transitions mostly (see Fig.~\ref{fig:EIT_setup}(b)). Both beams are blue detuned from the excited state by $\Delta = 120$ MHz. The intensity of the pump beam is tuned to create a light shift in the upper state of approximately the frequency of the motional mode to cool ($\sim 1.65$ MHz). The induced light shift is measured using the $6\textrm{S}_{1/2} \leftrightarrow 5\textrm{D}_{5/2}$, similar as described in Ref.~\cite{lechner2016electromagnetically}. The intensity of the probe beam is adjusted to be about 100 times weaker than the pump beam.

\subsection{A2: RF amplitude stabilization}\label{sec:RFstabilization}
\begin{figure}[h]
\centering
\includegraphics[width=1\columnwidth]{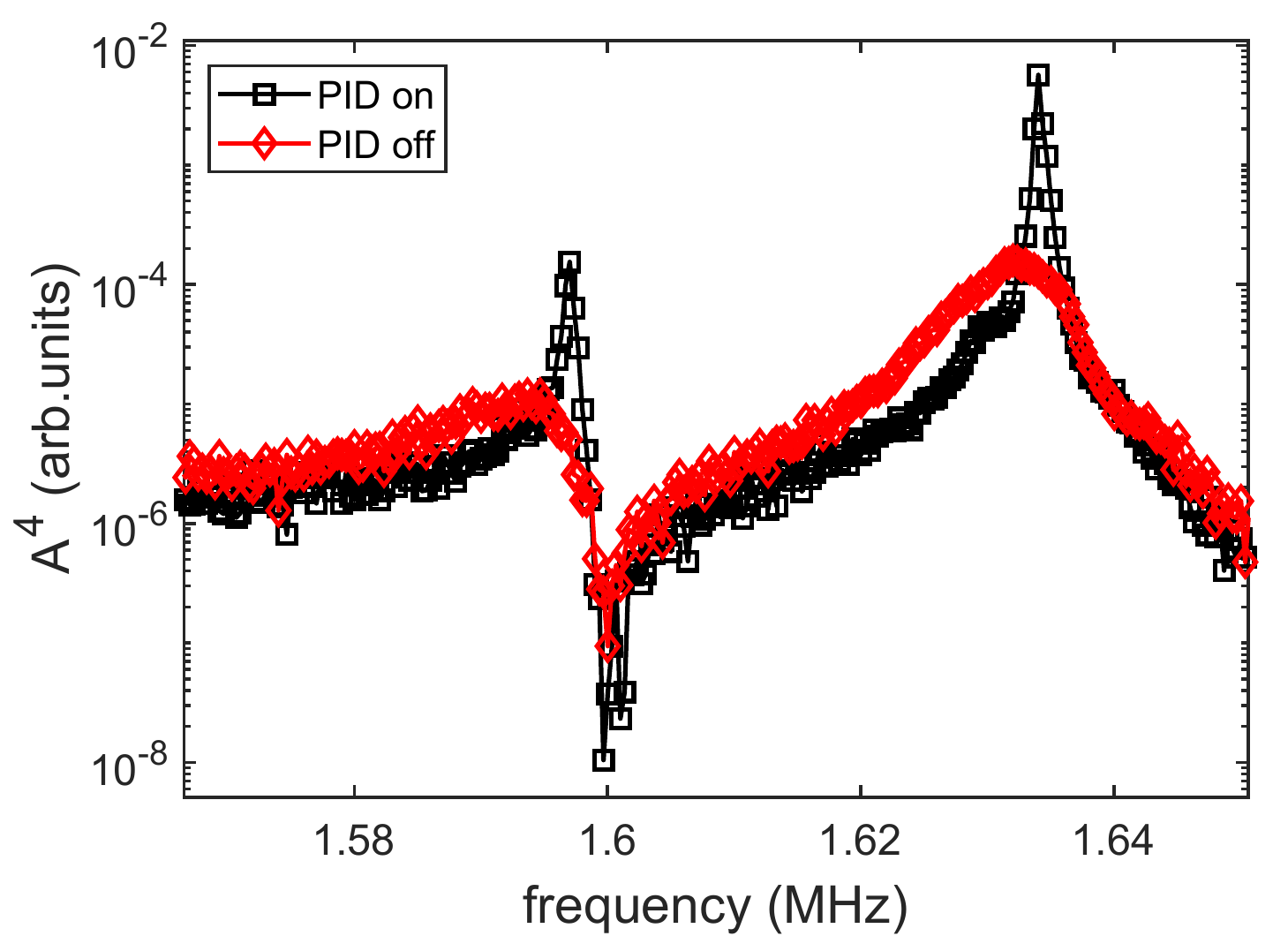}
\caption{Effect of the feedback stabilization for the amplitude of the trapping radiofrequency field. The two curves are acquired with and without the active feedback.}
\label{fig:PID_stabilization}
\end{figure}

The harmonic trapping pseudopotential of the linear Paul trap operated in the described experiments is generated by a radiofrequency field at $\Omega_\text{rf}/ 2\pi \sim$ $15.1 $~MHz. The stability of the radiofrequency is crucial to operate with a constant pseudopotential. Therefore, we stabilize the amplitude of the radiofrequency with an active feedback. At the trap electrodes, the radiofrequency is amplified via a resonant circuit. A coil assembled in the housing of the resonator picks up the signal directed towards the trap electrode. The electrical power probed by the pickup coil is averaged for a few milliseconds and used to correct the amplitude seeded to the resonant circuit via an analog PID circuit. Figure~\ref{fig:PID_stabilization} presents a measurement of the stabilization effect tested enabling or disabling the feedback activity. By deactivating the PID circuit stabilizing the amplitude of the rf trap drive, we observe a reduction by a factor of $\sim5-10$ in the coherence time of the oscillations. Thus, the active stabilization improves the stability of the pseudopotential, but it introduces a non-linearity which is visible on the low-frequency side of the oscillation peaks. In Fig.~\ref{fig:PID_stabilization} such non linearity can be seen in the data on the low frequency side of the peak located around $\sim 1.63$~MHz. This shape deformation is the cause of systematic uncertainties affecting the measurements reported in this Letter. Such uncertainties could be further reduced by using an electrode monitoring the amplitude instabilities closer to the ion position and by using an optimized PID circuit designed for this application. 

\subsection{A3: Interference fringe stability}\label{sec:fringestabilization}
In this Letter, we analyze the motion of the ion by interfering the atomic fluorescence with a mirror image. The stability of the interference pattern (Eq.~\ref{eq_fringe}) is fundamental to obtain consistent data during the acquisitions.

Without an active feedback, the mirror-ion distance drifts. To stabilize the distance, it is sufficient to integrate the count rate over $100-200$~ms to average away the effect of the ion oscillations. The rate ($\langle R \rangle = R_0 \sim 5\cdot10^4\,\text{s}^{-1}$) is then stabilized moving the mirror with a piezomechanical element. This feedback aims to correct for position drifts around the condition $\sin{\left(\frac{4\pi}{\lambda}q\right)}=0$, where maximum sensitivity to the ion oscillations is observed. In this way, the stability of the interference fringe is prolonged to $10^3$~s.

After this time, the experiment becomes sensitive to frequency drifts of the repumping laser driving the 6P$_{3/2}\leftrightarrow\text{5D}_{3/2}$ transition. These drifts modify the average emission rate of the ion. Since the interferometer is locked using the fluorescence of the atom, a variation of the average photon rate $\langle R \rangle$ changes the locking phase of interference, ultimately causing drifts on the detected motional spectral density. This limitation could be solved by interleaving atomic emission rate measurement and stabilization within the full experimental sequence or using an external absolute frequency reference as described in Ref.~\cite{Xie2019}.

A further and slower effect affecting the interference fringe is caused by misalignment of the optical setup which take place over several hours. We attribute to this slow drifts the great similarities of the curve having 1.0(3) and 0.5(2) mean phonon occupation number of Fig.~\ref{fig:EIT_cooling_amplitude}. As explained in the main text, the measurement was repeated five times to evaluate the uncertainty in the datapoints, thus requiring several hours of data taking.

\section{B: Ion oscillations from photon autocorrelation}
\subsection{B1: Physical model and photon autocorrelation}
The ion under continuous laser cooling evolves following free trajectories of oscillation. The trajectories are perturbed by events such as photon absorption and emission, collisions with the buffer gas or changes in the trapping electromagnetic field. All these events modify the ion trajectory reducing the coherence of the oscillations. They affect all the modes simultaneously. The system can be modelled in this way. The detected excursion $q(t)$ is the projection of the three modes along the optical axis of the confocal objective. Calling $w_j$ the director cosine of the optical axis with respect to the modes ($\sum w_j^2 = 1$), $A_j$ the oscillation amplitudes and $\phi_{j,k}$ random phase values, the signal can be modelled as
\begin{equation}\label{eq:motion_model}
    q\left(t\right)=\sum_{k}^{}\sum_{j}^{} A_j w_j \cos\left(\omega_j t + \phi_{j,k}\right) \, ,
\end{equation}
where $k$ is the number of phase jumps occurring into the acquisition.
The frequency response of each single mode can be calculated numerically on a discrete and finite support as
\begin{equation}\label{eq:transferfunction}
    L\left(\omega_j,\omega\right)=F\Bigg(\frac{\cos{\left(\omega_j t + \arctan\left(\frac{\gamma}{\omega_j}\right)\right)}e^{-\gamma t}}{\cos\left(\arctan\left(\frac{\gamma}{\omega_j}\right)\right)}\Bigg)
\end{equation}
where $F$ indicates the discrete Fourier transform operator and $\gamma$ the average rate of phase jumps. For large observation time compared to $1/\gamma$, the transfer function can be approximated by the formula for a dumped harmonic oscillator:
\begin{equation}\label{eq:Lorentziancontinuum}
    L\left(\omega_j,\omega\right)\sim\left(\omega_j^2-\omega^2-i\gamma\omega\right)^{-1} \; .
\end{equation}
The ion motion is analysed by computing the autocorrelation $g^{(2)}\left(t\right)$ of the photon events. For our analyses, we combined the photon sequences from the two APD detectors as it would be only one. It is a well known result of Fourier analysis that the autocorrelation function is the Fourier transform of the power spectrum. Considering only the linear term of the fringe expression (Eq.~\ref{eq_fringe}) the function $g^{(2)}$ can be calculated as:
\begin{equation}\label{eq:g2_to_spectrum}
    g^{(2)}\left(q,q,t\right) \propto F^{-1}\left|F q\left(t\right)\right|^2 \;.
\end{equation}
In this expression, the Fourier transform and its inverse should be computed on the entire available signal. For computational advantage dealing with rare photon events, it is more efficient to calculate $g^{(2)}$ as a histogram. We adopt histogram bins of $8-32$~ns size depending on the experiment. In each bin the coincidence photon events separated by a specific time delay are counted. The autocorrelation is computed from zero to $T = 2.1-4.3$~ms time delay. The truncation to $T$ aims to disregard events which are no longer correlated to each other because very far in time as compared to the oscillator coherence time $1/\gamma$. In practice, a large fraction of the possible bins is not considered by this analysis because the total acquisition time was several minutes. In this way, we can also limit the Fourier transform of the $g^{(2)}$ histogram on the available data to within $[0,T]$ without loosing important information. This approach boosts the computational speed by limiting both the histogram and the support of the Fourier transform. The operation is equivalent to analyzing the autocorrelation via a wavelet transform~\cite{Charles1992}. The wavelet analysis delivers a power spectrum averaged among neighbouring frequencies: a convolution in frequency space. The convolution smoothes the power spectral function and neighbouring frequencies become very similar in value. The analysis on the limited interval $[0,T]$ discards a lot of frequency components whose contribution is averaged to compute the remaining data points of the power spectrum. The final power spectrum can be written as
\begin{equation}\label{eq:powerpowerspectrum}
    p\left(\omega\right)\propto\left|F \left(\chi_{\left[0,T\right]}\left(t\right)g^{(2)}\left(q,q,t\right)\right)\right|^2 \; ,
\end{equation}
where $\chi_{\left[0,T\right]}\left(t\right)$ represents a square unitary pulse indicating the cut over the reduced interval $[0,T]$. As mentioned, we limited this Fourier transform $F$ to the support $[0,T]$. Thanks to the average mediated by the truncation, the result of this operation has the expression:
\begin{equation}\label{eq:modecoherent}
    p\left(\omega\right)\propto\left|\sum_{j}^{} A_j^2 w_j^2 L\left(\omega_j,\omega\right)\right|^2 \,.
\end{equation}
In the formula, if the approximation of Eq.~\ref{eq:Lorentziancontinuum} is not valid, the support $[0, T]$ should be used to evaluate the transfer functions $L\left(\omega_j,\omega\right)$ from Eq.~\ref{eq:transferfunction}. In this final power spectrum the modes appear interfering with each other.

\begin{figure}[h]
\centering
\includegraphics[width=1\columnwidth]{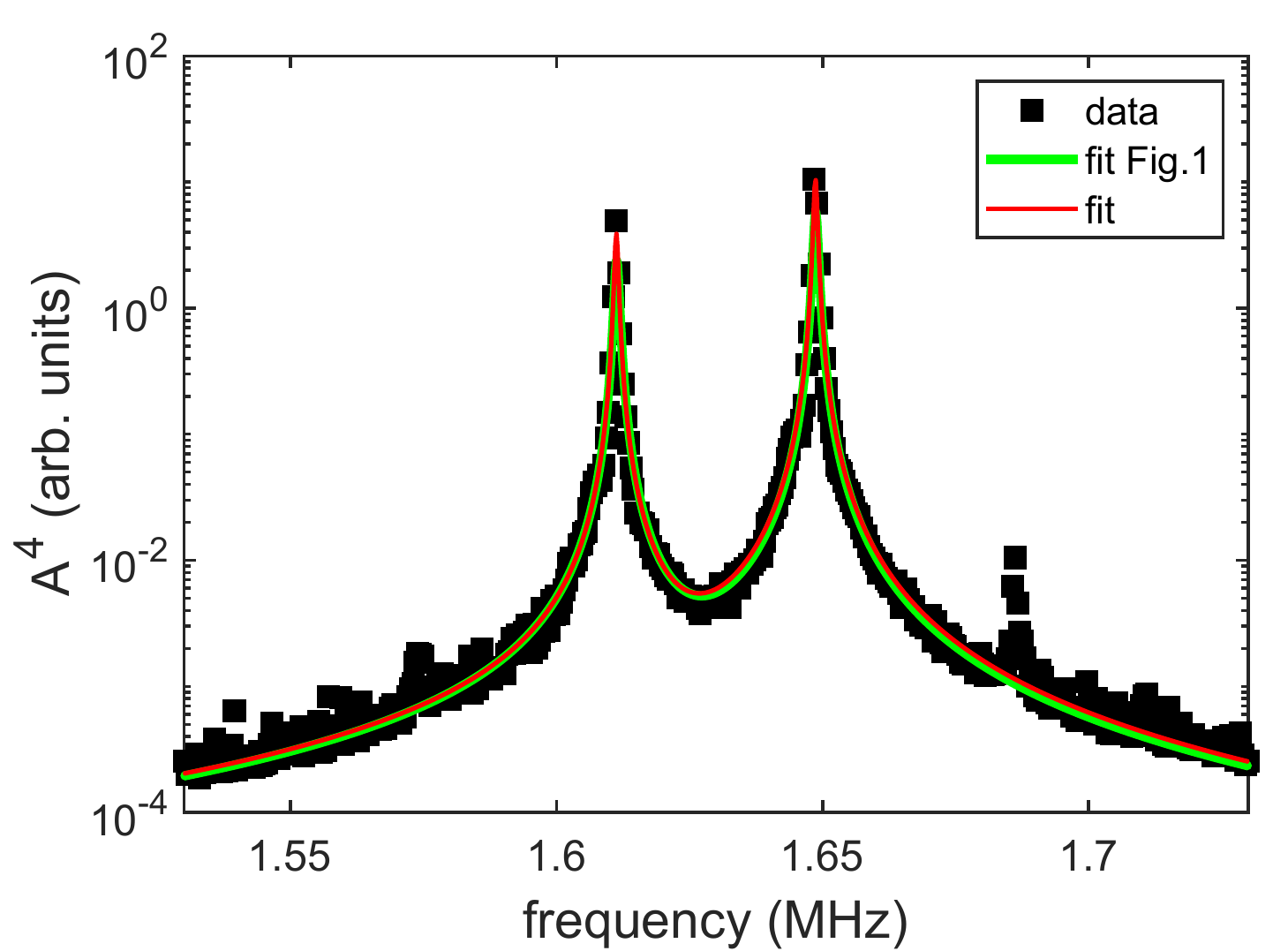}
\caption{Average of several power spectra obtained from the same autocorrelation function. The data are the same as analyzed in Fig.~\ref{fig:setup}. The curve \textit{fit} is a fit with Eq.~\ref{eq:modeincoherent}. The curve \textit{fit Fig.1} is also evaluated from Eq.~\ref{eq:modeincoherent}, but the coefficients are taken from the fit with Eq.~\ref{eq:modecoherent} on Fig.~\ref{fig:setup}(d).}
\label{fig:spectrumavg}
\end{figure}

In experiments involving photon counting and time-tagging, the autocorrelation analysis allows filtering the noise induced by the photon detection process. As a photon is detected, a single sharp peak is recorded in the signal (click), followed by antibunching. For this reason, the correlation overshoots at zero delay time because each photon is perfectly autocorrelating with itself and only sparsely in time with the others. Then, after time zero, no second photon can arrive before the dead time of the detector (or the atomic antibunching time) is elapsed. These noise contributions concentrate at the beginning of the autocorrelation. Thus, a selective filter can be applied by truncating the beginning of the correlation function: excluding the signal before the dead time $t_d$. In practice in Eq.~\ref{eq:powerpowerspectrum} the function $\chi_{\left[0,T\right]}$ is replaced with the shifted pulse $\chi_{\left[t_d,T+t_d\right]}$. This procedure introduces a time shift in all the cosinusoidal functions that reconstruct the autocorrelation. While the shift is suppressing the described noise contributions, it can modify the shape of the final spectrum. In our experiment, we confirmed that the modification was negligible for the final analysis by observing that with $t_d\sim100$~ns the truncation with $\chi_{\left[t_d,T+t_d\right]}$ or $\chi_{\left[2t_d,T+2t_d\right]}$ lead to compatible fit results. Thus, the data reported in Fig.~\ref{fig:setup} are analyzed with the truncation $\chi_{\left[t_d,T+t_d\right]}$. However, to overcome this limitation in other experiments, the average of several spectra with different cut pulses can be calculated. The average spectrum is computed from a group of consecutive discrete delays $t_m$ each generating a different cut pulse and, hence, a different spectrum. The delays $t_m$ should ideally cover a time span of the beating period of all the frequencies $\omega_j$. The modified spectrum can be calculated as:
\begin{align}
    p_s\left(\omega\right) &\propto \frac{1}{N}\sum_{m}^{}
    \left|F \left(e^{t_m\gamma}\chi_{\left[t_m+t_d,T+t_d+t_m\right]}\left(t\right)g^{(2)}\left(q,q,t\right)\right)\right|^2 \sim\\
    \label{eq:modeincoherent}
    &\propto\sum_{j}^{}\left| A_j^2 w_j^2 L\left(\omega_j,\omega\right)\right|^2 \; .
\end{align}
In Fig.~\ref{fig:spectrumavg}, a spectrum obtained via this analysis is presented. The measurement data are the same as analyzed in Fig.~\ref{fig:setup}. Here, the lineshapes of the oscillation peaks are symmetric and independent. Furthermore, the background noise is suppressed by the average over the different $t_m$.

Assuming energy equipartition among the modes, these analyses can be used to estimate the observation direction $w_j$. For example, disregarding the axial mode, the data presented in Fig.~\ref{fig:setup}(d) and in Fig.~\ref{fig:spectrumavg} confine the angle between our optical axis and the x-mode between 52$^\circ$ and 54$^\circ$. The uncertainty is dominated by systematic effects discussed in Suppl. Material A2.

In the experiments near the motional ground stated, the acquired photon stream is also analyzed via the autocorrelation function. In contrast to the previous analysis, here, photons are acquired during a short pulse (120~\si{\micro}s) of Doppler cooling. To reduce the effect of the short-time illumination, we divided the photon correlation function by a triangular function. The operation is equivalent to deconvolving the Fourier transform by the square pulse of illumination. Furthermore, before the Fourier analysis, the correlation was truncated at 105~\si{\micro}s to sample the oscillation frequency correctly to avoid aliasing and improving the numerical stability by reducing the deconvolution noise. 

\subsection{B2: Absolute amplitude measurements}\label{sec:highorder}
The instabilities of the interference fringe discussed in the Suppl. Material A3 prevent a reliable conversion from peak height in the spectrum to absolute amplitudes of oscillation. If an offset phase $\psi$ is added inside the sinusoidal function of the fringe equation (Eq.~\ref{eq_fringe}), the formula becomes
\begin{equation} \label{eq:fringeexpanded}
    R = R_0 \Bigg(1 + \mathcal{V} \sin\left(\frac{4\pi}{\lambda}q\right)\cos\left(\psi\right)+\mathcal{V}\cos\left(\frac{4\pi}{\lambda}q\right)\sin\left(\psi\right)\Bigg) \,.
\end{equation}
For slow drifts of $\psi$ and $R_0$ compared to the ion oscillation frequencies, the factors $R_0 \sin\left(\psi\right)$ and $R_0 \cos\left(\psi\right)$ can be time averaged independently from the terms containing $q\left(t\right)$. This approximation allows considering the equation valid over the entire photon stream with $R_0$, $\sin\left(\psi\right)$ and $\cos\left(\psi\right)$ replaced by their time average over the entire acquisition. Using this approximation, $R$ is further expanded in series of $q/\lambda$ for small excursion $q$. The series expansion in $q/\lambda$ contains odd and even terms which come from the expansion of $\sin\left(\frac{4\pi}{\lambda}q\right)$ and $\cos\left(\frac{4\pi}{\lambda}q\right)$ respectively. Each term in the expansion generates peaks at different frequencies in the spectrum. Odd and even terms can be grouped separately because of the different multiplicative coefficient in front. The phase $\psi$ is unknown, thus, the relative height of the peaks generated by the even and the odd terms are not determined. Conversely, all the peaks resulting by the odd terms, or alternatively the even terms, are proportional to the same multiplicative factor. Therefore, a comparison of peak heights within each group allows determining the absolute excursion of the oscillation with respect to the light wavelength. For our experiment, we selected the linear and third order peaks visible in Fig.~\ref{fig:spectrumavg}. Those third order peaks corresponds to the oscillation frequencies $2\omega_x - \omega_y\sim1.57$~MHz and $2\omega_y-\omega_x\sim1.69$~MHz. The relative amplitude of the linear and third order peaks indicates a peak-to-peak oscillation excursion in the radial modes of 115--125~nm (value reported in the text). The uncertainty of this value is dominated by the asymmetry of the lineshapes caused by the circuit that stabilizes the rf trapping voltage (see Suppl. Material A2). 

\section{C: Phase space analysis}
\subsection{C1: Frequency filter, phase space reconstruction and mean trajectory}\label{sec:filter}
\begin{figure}[h!]
    \centerline{\includegraphics[width=0.8
    \columnwidth]{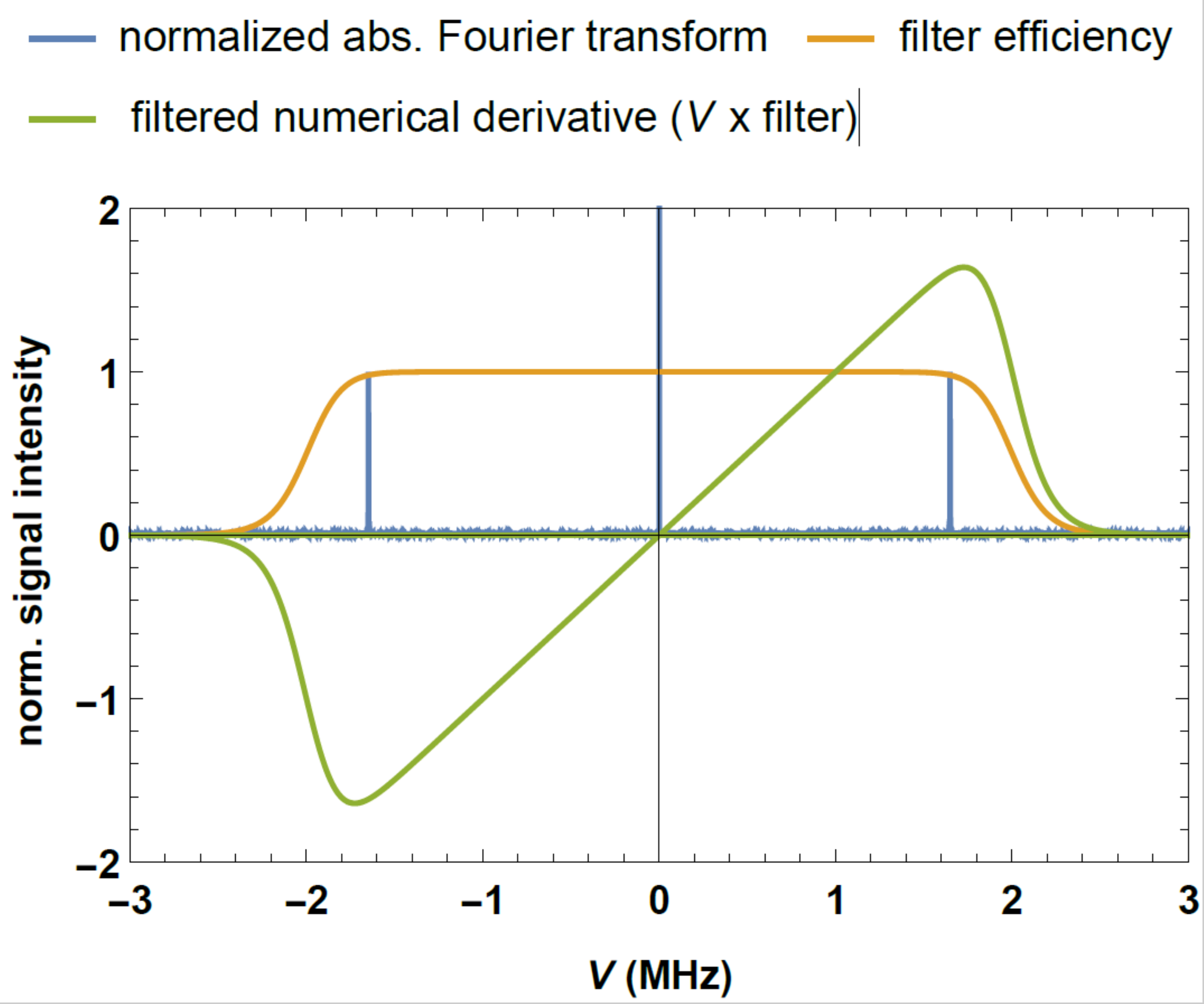}}
    \caption{Noise filtering for reconstruction of quadratures in the Fourier space. The blue curve shows a normalized absolute value of raw-data Fourier transform. The orange curve shows the applied filter as described in the text. The green line is the product of the derivative ramp by the low-pass filter.
    }
    \label{fig:filter_function}
\end{figure}
\begin{figure}
    \centerline{\includegraphics[width=0.8\columnwidth]{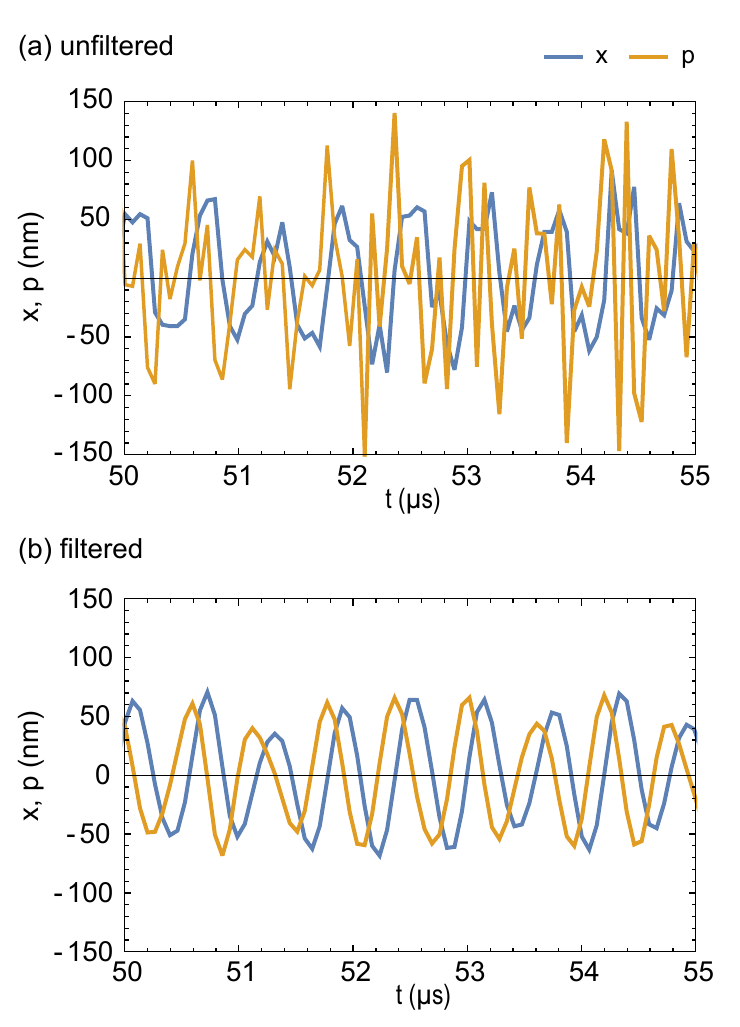}}
    \caption{Comparison of $x$ and $p$ quadratures calculated without and with filtering. }
    \label{fig:filtered_vs_unfiltered_XP}
\end{figure}

To reconstruct the trajectory of the ion in phase space, we synchronize time-tagging with the rf driving the oscillation. The drive is applied for 50~\si{\micro}s at the y-mode frequency for preferential excitation of that mode. As described in the text, we obtain a histogram of photon counts that is sensitive to the ion position with respect to time after averaging over several realizations. This constitutes the q-quadrature signal of the phase space. The second quadrature of the motion must be numerically reconstructed. The analysis is done considering that the momentum quadrature (p-quadrature) is proportional to the velocity of the ion inside the trap. The velocity can be calculated as the numerical time derivative of the position. We evaluated the numerical derivative in Fourier space by multiplying the transform of the q-quadrature by an imaginary linear ramp filter. In the reconstruction of the phase space trajectory, we took advantage of the motion characteristic to filter the noise at high frequencies. This was useful to reduce the noise in the reconstruction of the $p$-quadrature, since the derivative filter emphasizes the fluctuations in the high-frequency region of the spectrum. A linear low-pass filter was applied by multiplying our signal with a suitable function in Fourier space. In the analysis, we used a Butterworth filter of $x$-th order given by the expression $F(\mathcal{V}) = 1/[1 + (\nu /\nu_{c})^{l}]$, where $\nu_{c}$ is the cutoff frequency and $l$ is determining the cut-off sharpness for high frequencies. We used $\nu_{c} = 2$ MHz and $l = 20$ to retain all the frequency components of the secular motion. The spectrum of the raw data, the filter and the derivative ramp times the filter are plotted in Fig.~\ref{fig:filter_function}. The reconstructed quadratures with and without filtering are shown in Fig.~\ref{fig:filtered_vs_unfiltered_XP}. For experiments targeting particular oscillation frequencies, the filter can be applied only around the spectral range of interest.

After the analysis, we calculated the mean trajectory $x(t)$ by fitting an exponentially damped sine function of the form
\begin{equation}
 x(t) = A \sin (\omega t + \phi) \exp^{-\gamma t} + x_{\textrm{off}} 
\end{equation}
In the fit $A, \omega, \phi, x_{0}$ are the amplitude, frequency and phase of the oscillation, $x_{\textrm{off}}$ is an amplitude offset, and $\gamma$ represents exponential decay rate of the signal.

The results presented in Fig.~\ref{fig:phase} show the characterization of a particular state of motion following the dynamics of the experimentally driven mode. However, the technique can be used to characterize the full motion of the ion without restriction to a single secular mode. For example, in the presented data, a kink in the amplitude curve around 100--200~\si{\micro}s is compatible with a residual excitation of the x-mode by the drive.

\subsection{C2: Noise in the phase space reconstruction}

In the analysis of the histogram of the photon events, we noticed a slight asymmetry in the uncertainties between different positions corresponding to a high and low value of count rate of each bin. We attributed such asymmetry to shot noise deriving from the Poissonian distribution affecting the count rate for different absolute position of the emitter. Following this idea, when the ion emits more photons, we should observe larger fluctuations than at the opposite phase. In fact, the uncertainty at each signal level should be given by the square root of the average count value at that position. To prove our hypothesis, we analyzed the distribution of the residuals between the mean trajectory and the original data via a $\chi$-square test.

\begin{figure}[h]
    \centerline{\includegraphics[width=1.\columnwidth]{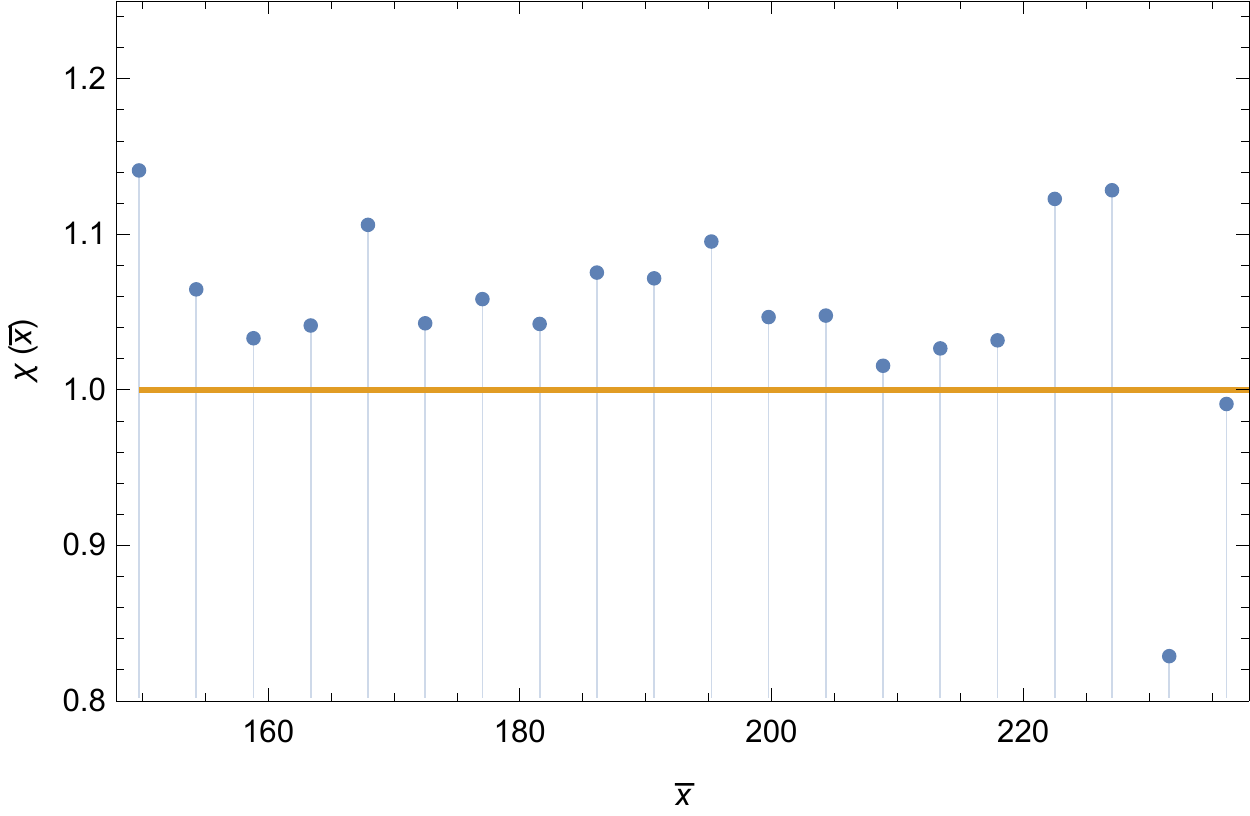}}
    \caption{Values of $\chi$ for each bin (blue points). The orange line show the expected value for purely shot noise.
    }
    \label{fig:chi_square_statistics}
\end{figure}

To do so, the mean trajectory data $x_i$ were binned according to their absolute position in 20 levels. Such subdivision corresponds to 20 equally spaced levels between a minimum of 150 counts per bin to a maximum of 237 counts per bin. We refer to the absolute position of the trajectory in count rate unit as $n_{j}$. For each bin $i$, the mean coincidence count $\overline{n_i}$ was estimated as the average point of the interval. For each level $\overline{n_i}$ we calculated the $\chi$-square test:
\begin{equation}
\chi(\overline{n}_{i}) = \frac{1}{N}\sum^{N-1}_{k=0}{\frac{\left(n_{k} - y_{k} \right)^2}{\overline{n}_{i}}},
\end{equation}
where $k$ runs only on the data of each subdivision and $y_k$ is the raw data count rate.
The values of $\chi_{i}$ are plotted in Fig.~\ref{fig:chi_square_statistics}. The curve is flat and the values for each interval bin are close to the ideal value one. The analysis indicates that most of the observed noise and its baseline can be modeled with shot noise.

\section{D: Nanoparticle application}\label{sec:noise}

A detailed analysis of how to apply the technique of this Letter to a dipolar scatterer is published in a different article. Here, we report the key elements necessary for a simplified description.
The detection sensitivity of the presented method can be analyzed thanks to the theoretical results for dipolar scatterer of Ref.~\cite{Tebbenjohanns2019}. The theory describes the detection of position by interfering the light scattered in a dipole pattern with a reference. In this work, an ideal setup to reach the Heisenberg limit is described. In the ideal case, the reference field has equal wavefronts as that of the dipole scattering particle and is located in the equilibrium position of the oscillations. Furthermore, to reach the ultimate sensitivity, the detection should cover the entire solid angle. Similarly, in the configuration adopted in the experiments here described, the atomic fluorescence of a dipole transition is self-referenced with an image of the field obtained by back reflection. The self-referencing ensures the compatibility of the wavefronts of the primary and reference fields. If a full solid angle realization were possible, a half sphere would be used for reflection and the opposite half for detection. In our scheme, the mirror generates an image of the oscillating ion which displaces symmetrically from the equilibrium position during oscillations. The fact that the both the ion and its image moves doubles the phase difference between the direct and the reflected fields at the detector as compared to a static reference. For this reason the configuration of our experiment is a practical implementation of the ideal theoretical arrangement because the information lost by covering half of the solid angle with a mirror, is recovered by a stronger intensity modulation.

\begin{figure}[h]
    \centerline{\includegraphics[width=1.\columnwidth]{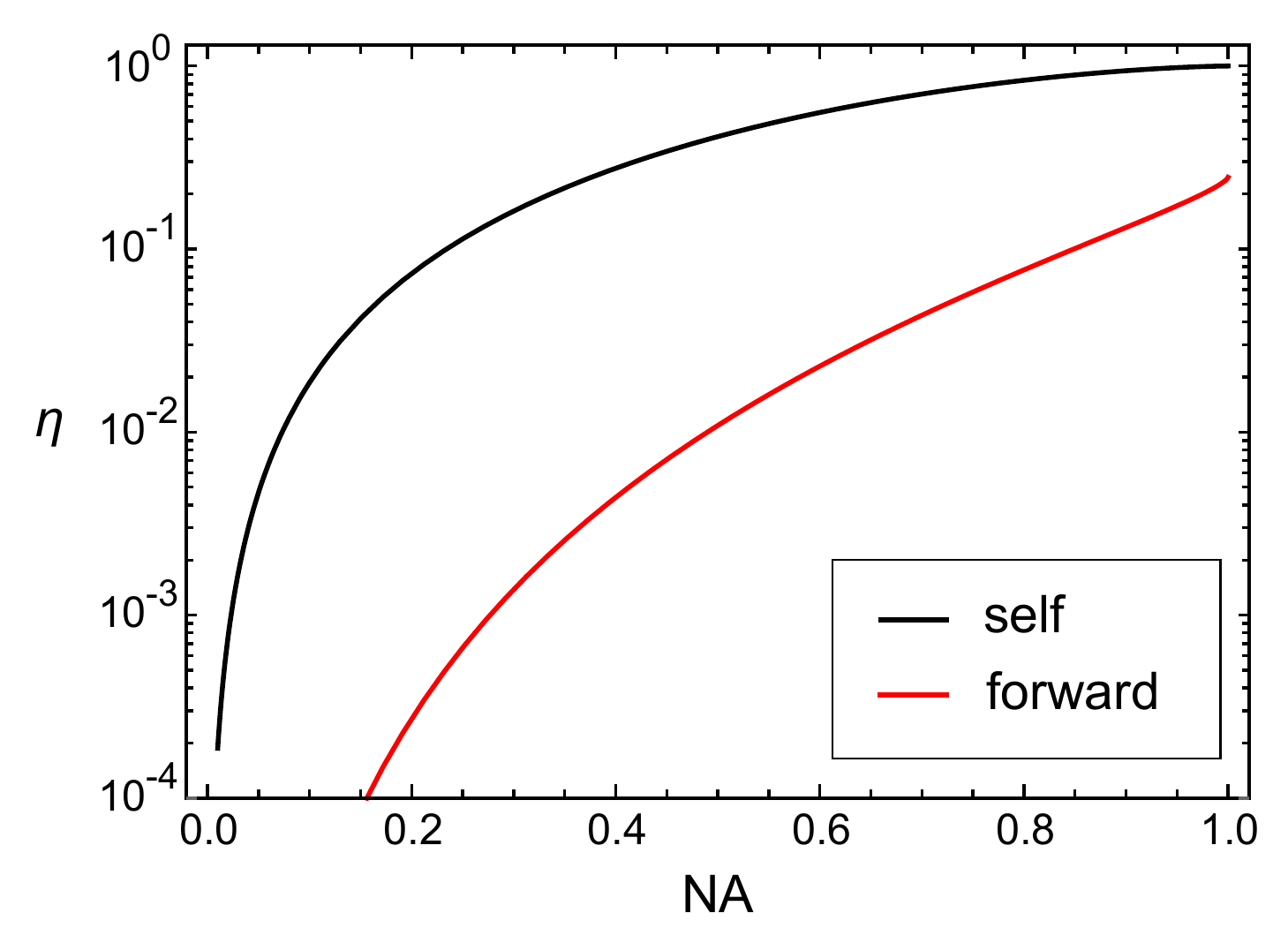}}
    \caption{Values of $\eta$ for the typical forward detection scheme (curve \textit{forward}) and the self-interference method described in this work (curve \textit{self}).
    }
    \label{fig:eta}
\end{figure}

In Ref.~\cite{Tebbenjohanns2019}, the authors define the efficiency for motion detection by comparing the measurable signal with the intrinsic measurement imprecision set by the back action. To model a realistic application, they assume the reference field to be the focused Gaussian beam which the nanoparticle scatters. Under this condition, the experiments must detect the interference along the beam direction. Similarly to our setup, the authors consider a setup composed by a confocal lens system. One lens focuses the light on the particle and the second collects the primary and scattered beam. With this assumptions the parameter $\eta$ is calculated: the ratio between the detection efficiency of a real-world measurement with limited NA and the ideal setup. $\eta$ is a number between 0, no detection, and 1, if the Heisenberg limit can be reached. We calculated the value of $\eta$ for our setup and compared it to the one provided for the forward detection scheme. Taking advantage of flexibility offered by our technique to be oriented along any direction, we aligned on three different orthogonal axis the linear polarization of the light illuminating the scatteres, the optical axis of forward detection scheme and the optical axis of self-homodyne method. $\eta$ is calculated for displacement of the ion along the optical axis of the self-homodyne technique. The comparison between the two values of $\eta$ as a function of NA is presented in Fig.~\ref{fig:eta}. Our setup delivers higher efficiency at lower NA because of the favorable orientation and, in addition, $\eta\rightarrow1$ for NA$\rightarrow1$. With a numerical aperture of NA$=0.4$, the predicted value of $\eta$ for forward detection described in Ref.~\cite{Tebbenjohanns2019} is $4.4\cdot10^{-3}$, while it could be as high as $0.28$ with our setup. For NA$=0.7$, $\eta$ would increase up to $4.3\cdot10^{-2}$ and $0.7$ respectively. 

\end{document}